\documentclass[twocolumn,aps,pra,amsmath,amssymb,superscriptaddress,floatfix]{revtex4-2}

\usepackage{color}
\usepackage{amstext}
\usepackage{amssymb}
\usepackage{graphicx}
\usepackage{xspace}
\usepackage{dcolumn}
\usepackage{bm}

\begin{document}

\title{Response of the Verwey transition in magnetite to a controlled\\ point-like disorder induced by 2.5 MeV electron irradiation}

\author{Ruslan Prozorov}
\email[Corresponding author: ]{prozorov@ameslab.gov}
\affiliation{Ames National Laboratory, Ames, Iowa 50011, USA}
\affiliation{Department of Physics \& Astronomy, Iowa State University, Ames,
Iowa 50011, USA}

\author{Makariy~A.~Tanatar}
\affiliation{Ames National Laboratory, Ames, Iowa 50011, USA}
\affiliation{Department of Physics \& Astronomy, Iowa State University, Ames,
Iowa 50011, USA}

\author{Erik~I.~Timmons}
\affiliation{Ames National Laboratory, Ames, Iowa 50011, USA}
\affiliation{Department of Physics \& Astronomy, Iowa State University, Ames,
Iowa 50011, USA}

\author{Marcin Ko\'{n}czykowski}
\affiliation{Laboratoire des Solides Irradi\'{e}s, \'{E}cole Polytechnique, CNRS, CEA, Institut Polytechnique de Paris, F-91128 Palaiseau, France}

\author{Tanya~Prozorov}
\affiliation{Ames National Laboratory, Ames, Iowa 50011, USA}
\affiliation{Department of Chemical and Biological Engineering, Iowa State University,
Ames, Iowa 50011, USA}

\date{14 May 2023}

\begin{abstract}
A controlled point-like disorder induced by low temperature 2.5 MeV electron irradiation
was used to probe the nature of the Verwey transition in magnetite,
$\textnormal{Fe}_{3}\textnormal{O}_{4}$. Two large single crystals, one
with optimal transition temperature, $T_{V}\approx121$ K, and another
with $T_{V}\approx109$ K, as well as biogenic nanocrystals, $T_{V}\approx110$ K, were examined.
Temperature-dependent resistivity is consistent with the semiconductor-to-semiconductor
sharp, step-like Verwey transition from a state with a small bandgap
of around 60 meV to a state with a larger bandgap of about 300 meV. The
irradiation causes an up-shift of the resistivity curves above the transition without transition smearing or broadening. It also causes an apparent down-shift of the resistivity maximum at high temperatures. In the lower $T_{V}$
crystal, the electron irradiation drives the transition temperature into a ``forbidden"
regime believed to separate the first order from the second order
phase transition. Contrary to this belief, the transition itself remains sharp and hysteretic without a significant change in the hysteresis width.
We conclude that the sudden change of
the bandgap accompanied by the monoclinic distortion and the change of
magnetic anisotropy is the reason for the Verwey transition in magnetite
and the effect of additional disorder is mostly in the smearing of the sharp gap edges near the Fermi level.
\end{abstract}
\maketitle

\section{Introduction}

The original studies of magnetite, $\textnormal{Fe}_{3}\textnormal{O}_{4}$,
have initiated one of the most fascinating topics in condensed matter
physics, the metal-insulator transition (MIT) \citep{Parks1926,Verwey1939,Verwey1941}.
(Although now we know that, specifically in magnetite, the transition
is a semiconductor-to-semiconductor type \citep{Liu2017}). It all
started in 1926 when an unexpected heat absorption at around 115 K -
117 K was found in heat capacity measurements by Parks and Kelley
\citep{Parks1926}. The authors write: ``The heat absorption, found
in the case of the magnetite crystals at about 115 K, constitutes
still another noteworthy feature. This temperature would seem to be
very low for a transformation in crystal structure in a metallic oxide,
the melting-point of which is above 1500 K. ... we think that the
observed heat effect may possibly be connected with a change in the
magnetic properties of the substance. ... Certainly this heat absorption
phenomenon should be investigated further.'' No graphics/figures
were shown, though in Ref.\citep{Parks1926}. Next, magnetic anomalies
at similar temperatures were reported \citep{Weiss1929,Li1932}, seemingly
agreeing with the assumption of the magnetic origin of the observed
behavior because no difference was found in x-ray diffraction measured
at room temperature and at liquid ``air'' (nitrogen) \citep{Li1932}.
Finally, thirteen years after the initial work, Verwey reported now
famous two orders of magnitude sharp increase of resistivity at around
120 K \citep{Verwey1939,Verwey1941}. Since that time, it became known
as the ``Verwey transition,'' and its microscopic origins have been
a subject of relentless experimental and theoretical research to this
day. Here we can only mention a very incomplete set of review
articles \citep{Bickford1953,Buckwald1975,Review1980,Honig1995,Muxworthy2000,Jackson2020,Friedrich2002,Garcia2004,Bazylinski2004,Rozenberg2006,Coe2012,Yu2014,BernalVillamil2015,Figuera2019}.

Stoichiometric magnetite is a mix of iron oxides, $\textnormal{Fe}_{3}\textnormal{O}_{4}=\textnormal{Fe}_{2}^{3+}\textnormal{O}_{3}^{2-}\cdot\textnormal{Fe}^{2+}\textnormal{O}^{2-}$,
which has an inverse spinel crystal structure with iron ions occupying
two distinct sites, $\textrm{Fe}_{3}\textrm{O}_{4}=\left[\textrm{Fe}^{3+}\textrm{(A)}+\textrm{Fe}^{3+}\textrm{Fe}^{2+}\textrm{\ensuremath{\left(\textrm{B}\right)}}\right]\cdot\textrm{O}_{4}$
where A sites are coordinated in a tetrahedron and B sites are coordinated
in an octahedron. Ionic magnetic moments at the A and B sites are
antiparallel, resulting in ferrimagnetism with an excess magnetic
moment of about $4\mu_{B}$ per formula units (f.u.).
For each formula unit, there are two B sites with spin $S=2.25\mu_{B}$
and one A site with $S=2.5\mu_{B}$. There are
8 f.u. in a deformed cubic cell with lattice constant of $a=8.4\:$\AA,
so each cubic unit cell contributes $32\mu_{B}$
of net moment. Magnetite's Curie temperature is about 800 K \citep{Aplesnin2007}.
Above the Verwey transition temperature, $T_{V}\approx121\:\textrm{K}$, the
B sites are charge-frustrated leading to a significant delocalization
of electrons, which results in a moderately conducting state with
the resistivity of about $0.005-0.01\:\Omega\cdot\mathrm{cm}$ at
room temperature and about $0.02\:\Omega\cdot\mathrm{cm}$ at $T_{V}$.
Verwey discovered that below $T_{V}$, the resistivity increases sharply
by about two orders of magnitude and continues to increase significantly
upon further cooling. He has explained this behavior in terms of charge
ordering in the B subsystem: Fe$^{2+}$ ions along the $\left[110\right]$
and Fe$^{3+}$ along $\left[\bar{110}\right]$ directions. This is
called a ``Verwey model'' \citep{Review1980,Verwey1939,Verwey1941}.
The resistive and thermodynamic signatures of the Verwey transition
are accompanied by the step-like change in sample magnetization and
all other thermodynamic, spectroscopic and transport properties \citep{Review1980}.
Similar to the resistive Verwey transition, the physics of the magnetic
anomaly apparently associated with it, is still debatable. It is known
that magnetic easy axis changes from the $\left[111\right]$ to $\left[100\right]$
 (within $0.2\:$\AA~ due to small monoclinic distortion) direction
below $T_{V}$. The $K_{1}$ term of magnetic anisotropy energy increases
by an order of magnitude, but the amplitude of the total magnetic
moment remains unchanged.

Since the original proposal, different mechanisms of the Verwey transition
considering new experimental results have been put forward. Resonant X-ray studies found no evidence of charge ordering and strong strong electron-phonon interaction was suggested as the cause of the Verwey transition \citep{Subias2004}. Anderson argued that in spinel's octahedral sites, nearest-neighbor Coulomb interactions can never lead to the long-range order, and even
the long-range Coulomb interaction are practically ineffective at least at higher temperatures \citep{Anderson1956a}. Based on the large thermopower and low conductivity observed in magnetite both below and above the Verwey transition, it was suggested that Anderson
localization due to randomness of electrons and impurities is involved
\citep{Mott2004,Whall1977,Whall1980}. In general, some form of electron localization
(e.g., trimerons \citep{trimerons1,Senn2015,trimerons2,trimerons3,trimerons4,Bosak2014}) is considered as one of the primary mechanisms of the transition. It was also demonstrated that
the collective Jahn-Teller distortion due to interacting local degeneracies
may play an important role \citep{Chakraverty1974,Pinto2006,Rigo1980,Senn2015}.
It supports some other ideas, such as the cooperative ordering of
molecular polarons below $T_{V}$ \citep{Yamada1980}. From another
angle, Mott suggested that the Verwey transition is a transition
from Wigner glass to Wigner crystal in an electronic subsystem, but there
was not much of experimental followup \citep{Mott1979,Mott1980}.
Considering non-metallic behavior, it was also suggested that the
condensation of active optical phonon mode could be the reason for
the transition \citep{Srivastava1983} and it seems that the oxygen-18
isotope effect study supported it \citep{Terukov1979}.

On the opposite side from electron localization mechanisms, there
is overwhelming experimental and theoretical evidence for the fairly
large ($\sim200$ meV) bandgap opening below the Verwey temperature
\citep{Banerjee2020,Ihle1980,Liu2017,Rozenberg2006,Yu2014}. This
gap could originate from the charge-orbital ordering due to the on-site
Coulomb interactions \citep{Jeng2004}. Yet, some details are not
quite clear. Some studies assert that the energy gap only opens below
$T_{V}$, while above the transition the electronic bandstructure
is gapless and involves molecular polarons hopping conductivity \citep{Ihle1980,Rozenberg2006,Yu2014}.
Scanning tunneling spectroscopy (STS) shows directly a clear smaller gap, of the order of 70 meV, above $T_{V}$ transition and a larger gap below \citep{Banerjee2020,Liu2017}. Diffusive x-ray scattering study finds short-range charge ordering above $T_V$ related to trimeron ordering below the transition. This short-range ordering persist at least up to room temperature and is, apparently, correlated with the Fermi surface nesting signatures \cite{Bosak2014}, which again would imply opening the small energy gaps at least on parts of the Fermi surface.

Another experimental puzzle in magnetite is the discontinuity of the
Verwey transition separating first order phase transition in relatively
clean compositions from the (alleged) second order transition in less
stoichiometric compounds. For example, in Fe$_{3\left(1-\delta\right)}$O$_{4}$,
the critical value of $\delta_{c}=0.0039$ \citep{Aragon1985,Kakol1989,Honig1986,Honig1995,Rozenberg1996,Shepherd1991}.
Ironically, the temperature-dependent resistivity in the latter regime
is highly hysteretic, pointing to a first-order nature of the transition.
No satisfactory explanation for this disparity exists \citep{Honig1986,Honig1995,Honig1990}.
Similar bifurcation was observed in chemically substituted magnetite,
$\text{Fe}_{3\left(1-x\right)}\text{M}_{3x}\text{O}_{4}$, with M
= Ni, Co, Mg, Al, Ti, Mn, Li (sorted from the lowest to the highest
rate of $T_{V}$ suppression) \citep{DjegaMariadassou1989,Kakol1992,Brabers1998,Gasparov2009}.
In this work, we show that the Verwey transition continues to
remain first order when driven to the ``forbidden'' range of $T_{V}$
values by disorder. It shows that off-stoichiometry and disorder are not equivalent.

The measurements under applied isostatic or directional (uniaxial strain) pressure support this assertion. Isostatic pressure suppresses
the Verwey temperature and, at around a critical value of $P_{c}\approx8$
GPa, recovers the higher-conductance ``metallic'' state, although
the details of $P_{c}$ and other features vary between different
studies \citep{Kakudate1979,Rozenberg1996,Mori2002,Rozenberg2006,Gasparov2012}.
Directional stress leads to the enhancement of $T_{V}$, most likely
due to favorable structural domain selectivity \citep{Aplesnin2007,Coe2012,Nagasawa2007}.
In either case, pressure/strain change $T_{V}$ continuously (but
at $P_{c}$ it disappears abruptly) and the significant hysteresis
in the measured quantity is observed, supporting our statement above.

Finally, another line of study was inspired by the reports of a significant
smearing and shifting of the Verwey transition in magnetite nanoparticles
of different size with initial reports showing the effect already
at about 100 nm, which is quite large for the local-moment magnetism
\citep{Goya2003,Lima2006,Mitra2014,Prozorov1999,Wang2004,Yang2004,Yu2014}.
This sub-field of magnetite nanoparticles grew very rapidly and, unfortunately,
it is very difficult or mostly impossible to know whether those numerous
reports have actually had nanocrystalline magnetite of a mix of iron
oxides or just maghemite, $\gamma-$Fe$_{2}$O$_{3}$. Many studies
rely on the conventional x-ray diffraction, but at room temperature
it is hard to distinguish between different oxides because both have
the same cubic structure and their lattice parameters are almost identical
\citep{Kim2012}. Moreover, it is likely that nanoparticles powder-like
samples actually contain different compositions in comparable proportions
\citep{Kim2012}. The only reliable feature, present only in magnetite,
the Verwey transition is smeared and broadened in nanoparticles assemblies
due to size and shape distributions, stress, strain and the variation
of stoichiometry, hence is difficult to identify. The biological magnetite
is the exception, as it comes as well-shaped particles of a similar
size \citep{Meldrum1993,Prozorov2007} and is proven to be Fe$_{3}$O$_{4}$
by the magnetic induction mapping using electron holography \citep{DuninBorkowski1998,Simpson2005}
and NV-centers in diamond \citep{Sage2013}). One of the most detailed
systematic investigations where a great deal of attention was devoted
to the identification of magnetite as the primary phase, expectantly
showed that the Verwey transition remains practically unchanged down
to a nanoparticle size of 8 nm, below which $T_{V}$ drops very rapidly
\citep{Lee2015}. Similar conclusions were drawn by considering biological
magnetite of magneto-tactic bacteria where the chemical composition is
preserved by the lipid membranes preventing oxidation \citep{Prozorov2007}.
Of course, an additional technical problem is the size-dependent superparamagnetic
nature of magnetite nanoparticles \citep{Prozorov1999,Prozorov2004,Prozorov2007,Prozorov2019}.
The magnetic signature of the Verwey transition measured in nanoparticle
assemblies is smeared simply because of the size-dependent blocking
temperature, $T_{B}$, which quickly becomes lower than bulk $T_{V}$
rendering magnetic signature of the Verwey transition ill-defined,
while direct transport measurements on single nanoparticles are not
generally possible in most labs. In case of fairly large (50 nm) biological
nanoparticles, the Verwey transition is somewhat smeared, yet, it
is quite well-defined because blocking temperature is close to room
temperature \citep{Prozorov2007} (also, see the data below).

With so much experimental and theoretical effort devoted to magnetite
and the Verwey transition over almost 100 years, what new can one
possibly do and find? In order to shed some more light on this fascinating
topic, we used a controlled point-like disorder as a probe. We stress
that chemical doping or intentional variation of Fe/O ratio conducted
in many works in the past, is not equivalent to the artificial uncorrelated
disorder, because such disorder does not ``dope'' the system and
does not change the Fermi energy. (This was checked using Hall effect
measurements on pristine and disordered samples, albeit in a different
system \citep{Prozorov2019}.) If Verwey's charge ordering, Anderson
localization, or the formation of Wigner crystal scenarios were realized,
we would expect to observe a substantial smearing of the transition upon the introduction of random disorder. In the case of polarons, it is unclear
what to expect with the added disorder. If, on the other hand, the disorder
only affects the bandwidth and/or introduces the extra ``impurity''
band, the transition temperature, $T_{V}$, would shift to the lower
temperature, because the clean band-gap will become smaller but the
transition itself would remain sharp and hysteretic. Unfortunately,
we did not find systematic theoretical studies of the possible effects
of point-like disorder on the Verwey transition. It was suggested
that impurities may lead to closing up the gap above $T_{V}$ \citep{Brabers1999b,Brabers1998}.
However, this is not supported by the experiment and the Authors considered
an actual ionic substitution rather than dilute scattering centers.
Perhaps our work will serve as the motivation for further studies.

\begin{figure}[tb]
\includegraphics[width=8cm]{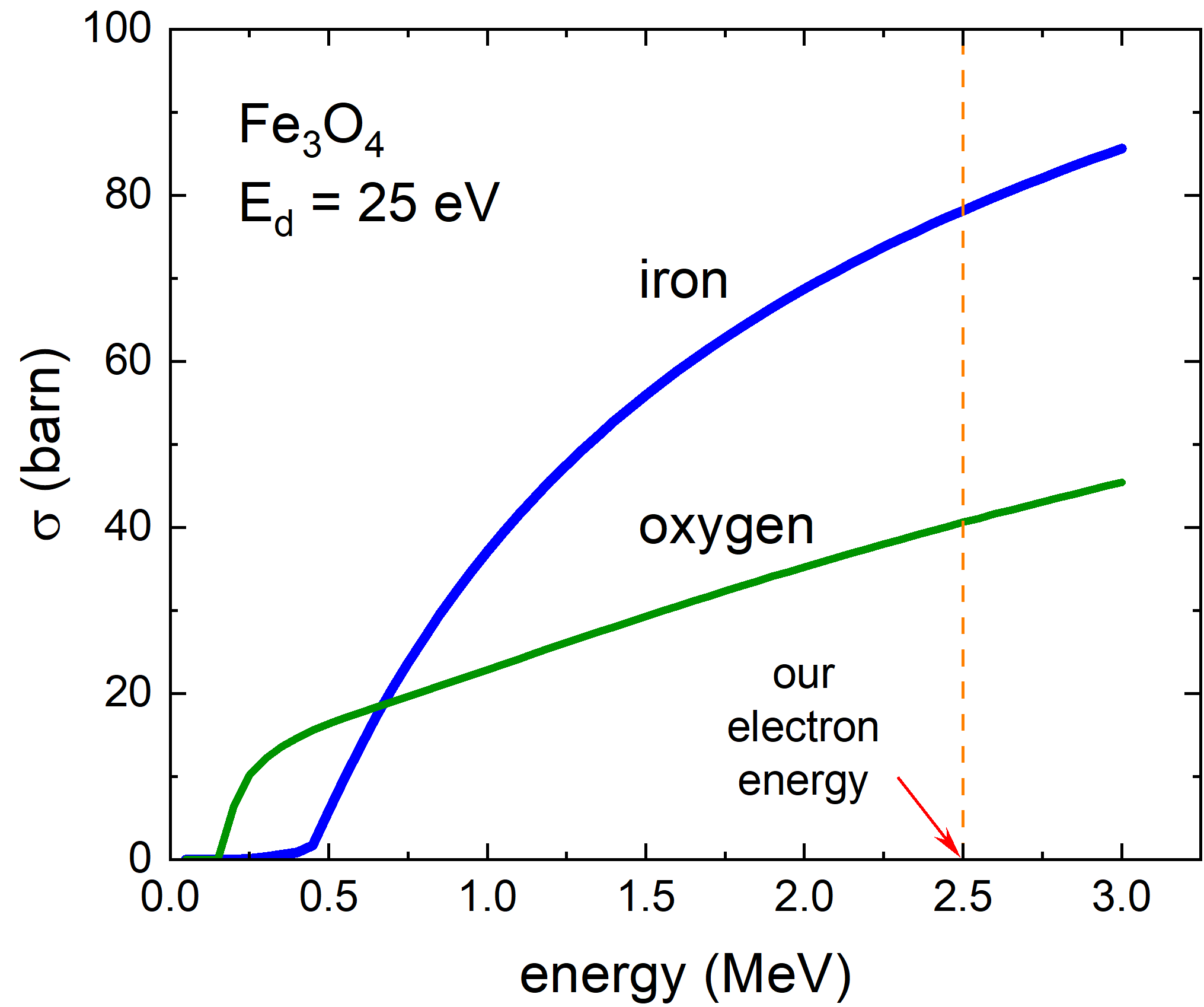}
\caption{Partial cross-sections of knock-out defect creation (both, vacancies and interstitials) for iron (blue line) and oxygen (green line) ions in stoichiometric magnetite, $\textrm{Fe}_{3}\textrm{O}_{4}$, as a function of electron energy assuming the typical displacement energy threshold, $E_{d}=25\:\mathrm{eV}$. At the operating energy of 2.5 MeV, the total cross-section is $\sigma=56.7$~barn, which leads to the estimate of $2.48\times10^{-3}$ defects per 1 C/cm$^{2}$ per formula unit ($dpf$), $\textnormal{Fe}_{3}\textnormal{O}_{4}$.}
\label{fig1:SECTE}
\end{figure}

\section{Experimental}

\subsection{Controlled artificial point-like disorder}

The point-like disorder was introduced at the SIRIUS facility in the Laboratoire des Solides Irradi\'{e}s at \'{E}cole Polytechnique, Palaiseau, France. Electrons were accelerated using a pelletron-type linear accelerator to the energy of 2.5 MeV, reaching 0.985 of the speed of light \citep{RullierAlbenque2000}. Such electrons are capable of knocking out ions from crystal lattice creating the so-called Frenkel pairs of vacancy-interstitial \cite{Damask1963,Thompson1969,PRX,KChoSUST,KChoNbSe2}.

Figure \ref{fig1:SECTE} shows the ion-type-resolved cross-sections
of the defects creation calculated using SECTE (``Sections Efficaces
Calcul Transport d'\'{E}lectrons") software, developed at \'{E}cole Polytechnique
(Palaiseau, France) by members of the ``Laboratoire des Solides Irradi\'{e}s",
specifically for the interpretation of MeV-range electron irradiation
using their pelletron-type linear accelerator, SIRIUS \cite{SIRIUS}.
Using atomic weights averages, the SECTE software interpolates the ion knock-out cross-sections tabulated by O. S. Oen \citep{Oen1973}. The input parameters are the chemical composition of the substance of interest, the direct head-on knockout energy, $E_{d}=25$~eV, and the energy of the projectiles (electrons), 2.5 MeV in our case. We used the commonly assumed
value of $E_{d}=25$~eV  \citep{Damask1963,Thompson1969} since the exact numbers are unimportant because we are only interested in the order-of-magnitude estimates. We obtained the partial cross-sections, 40.6 barn for oxygen and 78.2 barn for iron. Therefore, the
total cross-section is 56.7 barn. With the total beam current, measured behind the sample using a Faraday cage, $\mathrm{2.7\:\mu A}$ through a 5 mm in diameter circular diaphragm,
the electron beam flux is $8.6\times10^{13}$ electrons/($\mathrm{s}\cdot\mathrm{cm^{2}}$).
The acquired irradiation dose is conveniently measured in C/cm$^{2}$,
where 1 C/cm$^{2}$ = 6.24 $\times$ 10$^{18}$ electrons/cm$^{2}$.
This gives $3.54\times10^{-4}$ defects per 1 C/cm$^{2}$ per atom
(dpa), or $2.48\times10^{-3}$ defects per 1 C/cm$^{2}$ per formula
unit ($dpf$), $\textnormal{Fe}_{3}\textnormal{O}_{4}$ (not conventional unit cell with eight formula units!). An interesting feature of Fig.\ref{fig1:SECTE} is that while at 2.5 MeV iron defects dominate, if the energy is reduced to around 0.5 MeV, the only defects produced will be oxygen interstitials and vacancies. Therefore, it is possible to conduct ion-specific study to determine what kind of defects affect the properties the most.

For the doses used in this work of 2, 3.8, 4.1, and 6.3 C/cm$^{2}$
we estimate the concentration of defects to be 5, 9, 10 and 16 defects
per thousand of formula units, or 40, 75, 81, 125 defects per thousand
of conventional deformed cubic unit cell, or 0.5, 0.9, 1.0, 1.5 $\times$
10$^{6}$ defects/cm$^{3}$. These estimates are made under the assumption
that all collisions are head-on and there is no recombination of the
Frenkel pairs. During electron irradiation, the sample is immersed
in liquid hydrogen at around 20 K to remove significant heat of collisions
and prevent immediate recombination and clusterization of the defects.
On warming to room temperature, the induced defects partially anneal
with some Frenkel pairs recombine and ions in interstitial positions
migrate to various sinks (dislocations, surfaces etc.) leaving behind
a quasi-equilibrium distribution of vacancies which have much higher
barrier for diffusion. The resultant level of the induced disorder
is reflected in the change of sample resistivity at room temperature
where carrier density is roughly constant, and the only change comes
from the differences in the residual resistivity before and after
irradiation. Typically about 30\%-50\% of the defects are lost as
judged by the in-situ resistivity measurements \cite{PRX,KChoSUST}.
For magnetite, such measurement is difficult due to its gapped semiconducting
nature.

For comparison, 1\% $\left(\delta=0.01\right)$ of iron vacancies,
$\textnormal{Fe}_{3\left(1-\delta\right)}\textnormal{O}_{4}$, found
in biogenic magnetite \citep{Jackson2020} and synthetic magnetite
with different synthesis/annealing protocols \citep{Aragon1985,Honig1986,Kakol1989},
or ionic substitutions $\text{Fe}_{3\left(1-x\right)}\text{M}_{3x}\text{O}_{4}$,
substitutions with M=Ni, Co, Mg, Al, Ti, Mn, Li (sorted from lowest
to highest rate of $T_{V}$ suppression) \citep{DjegaMariadassou1989,Kakol1992,Brabers1998,Gasparov2009}
corresponds to $dpf=3\delta/7=4.3$ defects per 1000 formula units,
comparable with the concentration of defects induced by electron irradiation
used in this paper. An important advantage of controlled irradiation
is that the same physical sample is studied before and after irradiation
without taking off the contacts for transport measurements. This removes
any ambiguity that exists when comparing different samples that were
subject to different annealing protocols or chemical substitutions.

\begin{figure}[tb]
\includegraphics[width=8cm]{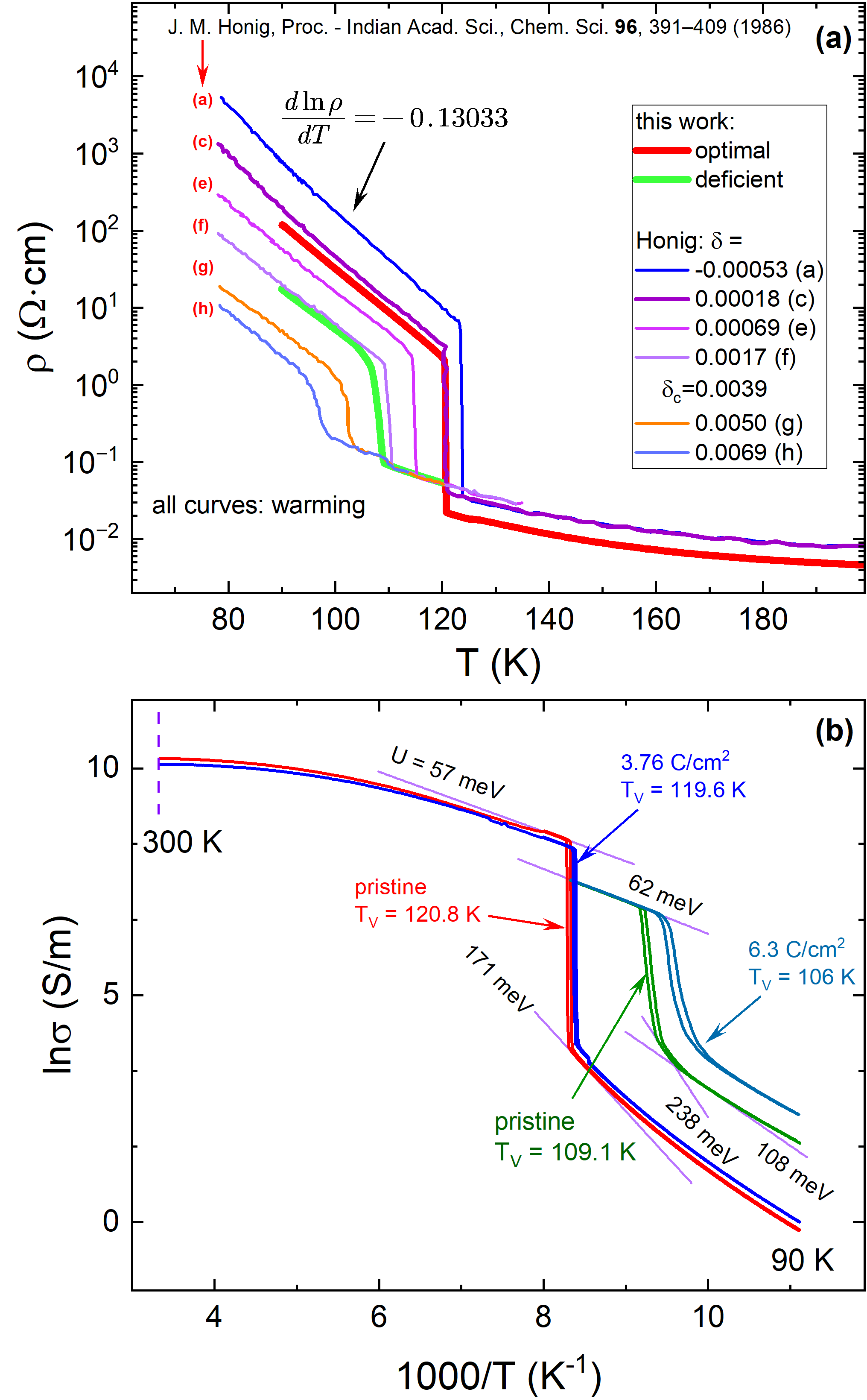}
\caption{
(a) Temperature-dependent resistivity on a $\log{}_{10}$ scale for the samples used in this study, solid red line - optimal, solid green line - oxygen deficient. For reference  we plot the data after Honig \cite{Honig1986} for oxygen-deficient samples. (b) Arrhenius plots of conductivity, $\ln{\sigma}$ vs. $1000/T$, for the samples before and after irradiation.}
\label{fig:lnR(T)HONIG}
\end{figure}

\subsection{Samples and experimental techniques}

Single crystals of $\textnormal{Fe}_{3}\textnormal{O}_{4}$ were obtained from J. M. Honig, which provided an excellent opportunity to directly compare our measurements before electron irradiation with his published data that were the first published systematic electrical resistivity study of  $\textnormal{Fe}_{3-\delta}\textnormal{O}_{4}$ on a set of samples with different $\delta$. The crystals grown by using the so-called skull melting technique described elsewhere
 \citep{Harrison1983,Honig1986,Honig1995}. Samples for transport
measurements with typical sample sizes of (1-2)$\times$0.3$\times$0.1
mm$^{3}$ (length$\times$width$\times$thickness) were cut and polished
from large single crystals. The short side corresponded to {[}011{]}
direction.

Electrical transport measurements were performed in a \textit{Quantum Design} physical property measurement system (PPMS) using a standard four-probe method with electrical current flowing along the longer side. The contacts were
made by soldering silver wires with indium solder and the follow-up mechanical strengthening with conductive Dupon 4929N silver paste \cite{TanatarFeSe2016}. The contact resistance of these contacts was below 100 $\mu\Omega$, and
they were mechanically stable to withstand electron irradiation \cite{PdTeeirr} .
Importantly, the contacts remained intact between measurements and
irradiation runs, thus enabling quantitative characterization of the resistance change without invoking changes in the sample geometry.

Magnetite nanocrystals, approximately $20\times20\times50$ nm in
size, where extracted from the lysed cells of magnetotactic bacteria,
MV-1 strain \citep{Bazylinski2004,Bazylinski1988,Prozorov2007}. Details
of the extraction process and magnetic characterization of different
biogenic magnetite strains and single crystals are published elsewhere
\citep{Prozorov2007}. The retrieved nanoparticles were smeared on
a standard carbon grid for transmission electron microscopy (TEM) studies.
For irradiation, the grids were enclosed in thin aluminum pouches
for protection. Magnetization measurements were carried out using
a \emph{Quantum Design} magnetic property measurement system (MPMS).
Magnetite nanocrystals were measured directly before and after irradiation
in their aluminum enclosures.

\section{Results}

We first look at the stoichiometric single crystal and then compare
the results with thouse for off-stoichiometric crystal. The samples for electron irradiation
and measurements were cut and shaped from these large crystals as
described in the Experimental section. The close to optimal composition
crystal had $T_{V}\approx121$ K \citep{Honig1995,Honig1990}, while
another had substantially lower $T_{V}\approx109$ K, it showed equally
sharp transition and thermal hysteresis in $\rho\left(T\right)$. For completeness,
we also study the effect of disorder on biological magnetite extracted
from magnetotactic bacteria \citep{Blakemore1975,Frankel1979,Bazylinski1988,Simpson2005,Prozorov2007,Jackson2020}.

\begin{figure}[tb]
\includegraphics[width=8cm]{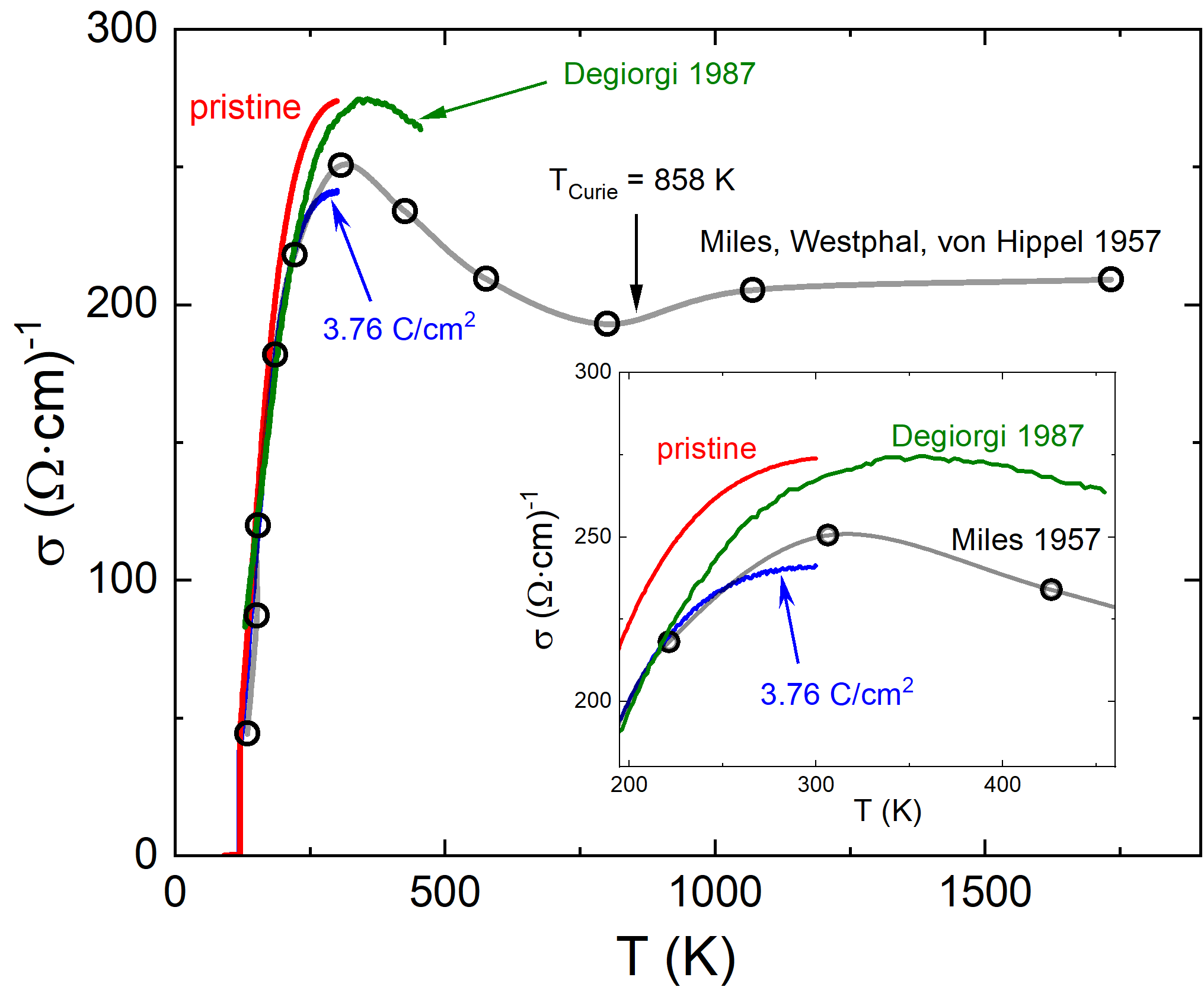}
\caption{Electrical conductivity in $(\Omega \cdot \textrm{cm})^{-1}$ of our pristine magnetite crystal (solid red curve) and irradiated with the dose of 3.76 C/cm$^2$ (solid blue curve), compared to the literature data up to almost 500 K in Ref.~\cite{Degiorgi1987} (solid green curve), and extended to much higher temperatures, well above the Curie temperature of 858 K, from Ref.~\cite{Miles1957}, shown by black open circles connected by a gray spline line. The inset focuses on a region of conductivity maximum occurring at around 300 - 350 K. }
\label{fig:high-T}
\end{figure}

Figure \ref{fig:lnR(T)HONIG}(a) shows temperature dependent resistivity
on a $\log_{10}$ scale versus temperature for two magnetite single
crystals used in this study, with $T_{V}\approx121$ K and $T_{V}\approx109$
K. For comparison, we re-plot Honig's data from his Fig.13 \citep{Honig1986},
which was the first systematic reported study of this kind. Remarkably,
both below and above $T_{V}$ , the slopes $d\lg\rho/dT$ are practically
identical between all these samples despite the fact that Honig has
grouped his samples into groups A (1st order transition) and B (2nd
order transition.). In order to examine the thermally activated population
of carriers in the conduction band, $n_{c}$, Fig.~\ref{fig:lnR(T)HONIG}(b)
presents the natural logarithm of electrical conductivity, $\sigma\sim n_{c}\sim$exp$\left(-U/T\right)$ plotted versus $1000/T$. The linear fits around the transition are
shown by solid lines in Fig.~\ref{fig:lnR(T)HONIG}(b). The activation energy,
$U$, is estimated to be about 60 meV above $T_{V}$, and about 200
meV just below, regardless of the $T_{V}$ value, and in a fair agreement
with the scanning tunneling spectroscopy (STS) data \citep{Banerjee2020}.
In temperature equivalent, these gaps correspond to about 700 K above
the transition and 2700 K just below $T_{V}$. Above the transition this activation energy corresponds to an energy gap in band-gap model  \cite{Cullen1971}, polaron activation energy in the polaron transport model \cite{polaron}, and to energy distance to mobility edge in weak localization model \cite{Whall1977}.

In Fig.~\ref{fig:high-T} we compare temperature-dependent conductivity $\sigma$ of high $T_V$ sample (red curve) with previous measurements over extended temperature range, by DeGiorgi \cite{Degiorgi1987} and by Miles et al.  \cite{Miles1957}. Similar to these studies, conductivity tends to saturate before reaching a peak at $\sim$350~K. A metallic-like decrease of conductivity on heating above the peak may suggest a cross-over to an intrinsic conductivity regime, supporting band transport model \cite{Cullen1971}.
Irradiation with 3.76 C/cm$^2$ (blue curve), leads to a more pronounced flattening of $\sigma(T)$ curve, suggesting a shift of the maximum to lower temperatures. This may be natural for a semiconductor with higher carrier density, as expected for crystals with higher defect density after irradiation.

\begin{figure}[tb]
\includegraphics[width=8cm]{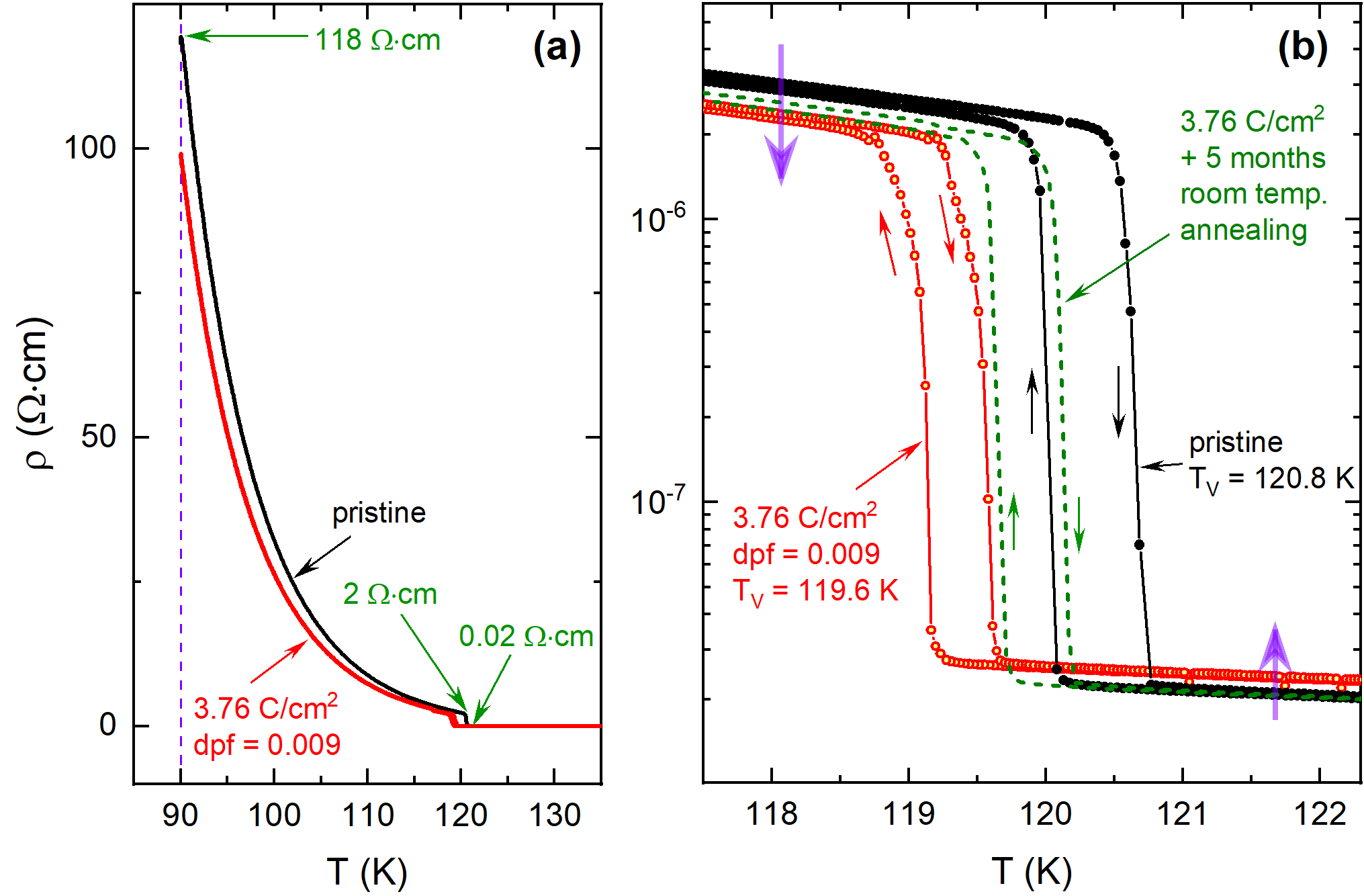}
\centering
\caption{(a) Temperature-dependent resistivity of optimal single crystal before
(black line) and after (red line) electron irradiation with a substantial
dose of 3.76 C/cm$^{2}$ = 2.3 $\times$ 10$^{19}$ electrons/cm$^{2}$
producing a maximum of 9 defects per 1000 formula units. There is
a clear signature of the Verwey transition at around 121 K where the
resistivity changes by two orders of magnitude from $0.02\:\mathrm{\Omega}\cdot\mathrm{cm}$
above $T_{V}$ to $2\:\mathrm{\Omega}\cdot\mathrm{cm}$ below. (b)
Resistivity plotted on a logarithmic scale zooming in on the Verwey
transition region revealing substantial hysteresis signifying a strong
1st order phase transition. In addition to the curves shown on the
left panel, here green dashed line shows the evolution of the Vervey
transition after five months of passive room-temperature annealing
on the shelf. Importantly, all curves remain parallel at the transition
showing no broadening. Also note that irradiation causes an increase
of $\rho$ avove $T_{V},$ and a decrease below as indicated by violet
arrows.}
\label{fig:R(T)optimal}
\end{figure}

Figure \ref{fig:R(T)optimal} presents electrical transport measurements
in magnetite single crystal of optimal composition. The black curve
is the pristine sample and red is the measurement of a sample after
a substantial dose of electron irradiation, 3.76 C/cm$^{2}$ = 2.3
$\times$ 10$^{19}$ electrons/cm$^{2}$ producing a maximum of 9
defects per 1000 formula units. Panel (a) shows $\rho\left(T\right)$
on the linear scales. There is a clear signature of the Verwey transition
at around 121 K and, as it was first observed by Verwey \citep{Verwey1939},
the resistivity changes by the two orders of magnitude from $0.02\:\mathrm{\Omega}\cdot\mathrm{cm}$
above $T_{V}$ to $2\:\mathrm{\Omega}\cdot\mathrm{cm}$ below. At
90 K, resistivity grows exponentially to $198\:\mathrm{\Omega}\cdot\mathrm{cm}$.
Analysis shown in Fig.\ref{fig:lnR(T)HONIG}, gives the energy gap
for for carriers in the conduction band, $U=171$ meV below $T_{V}$
and $57$ meV above. Panel (b) shows the resistivity plotted on a
logarithmic scale zooming in on the Verwey transition region, revealing
substantial hysteresis between cooling with its transition temperature,
$T_{V}^{c}$, and warming with, $T_{V}^{w}>T_{V}^{c}$, signaling
a strong 1st order phase transition. (Throughout the paper we use
higher value, $T_{V}^{w}$, speaking of the generic transition, $T_{V}.$)
Remarkably, the transition temperature in the irradiated sample shifts
by a significant 2.2 K ($\sim2\%$), but does not smear. All curves
remain practically parallel at the transition showing no broadening
or smearing. If anything, the hysteresis becomes a little bit smaller.
In addition to the curves shown in panel(a), the green dashed line
in panel (b) shows the evolution of the Verwey transition after five
months of passive room-temperature annealing on the shelf.

\begin{figure}[tb]
\includegraphics[width=8cm]{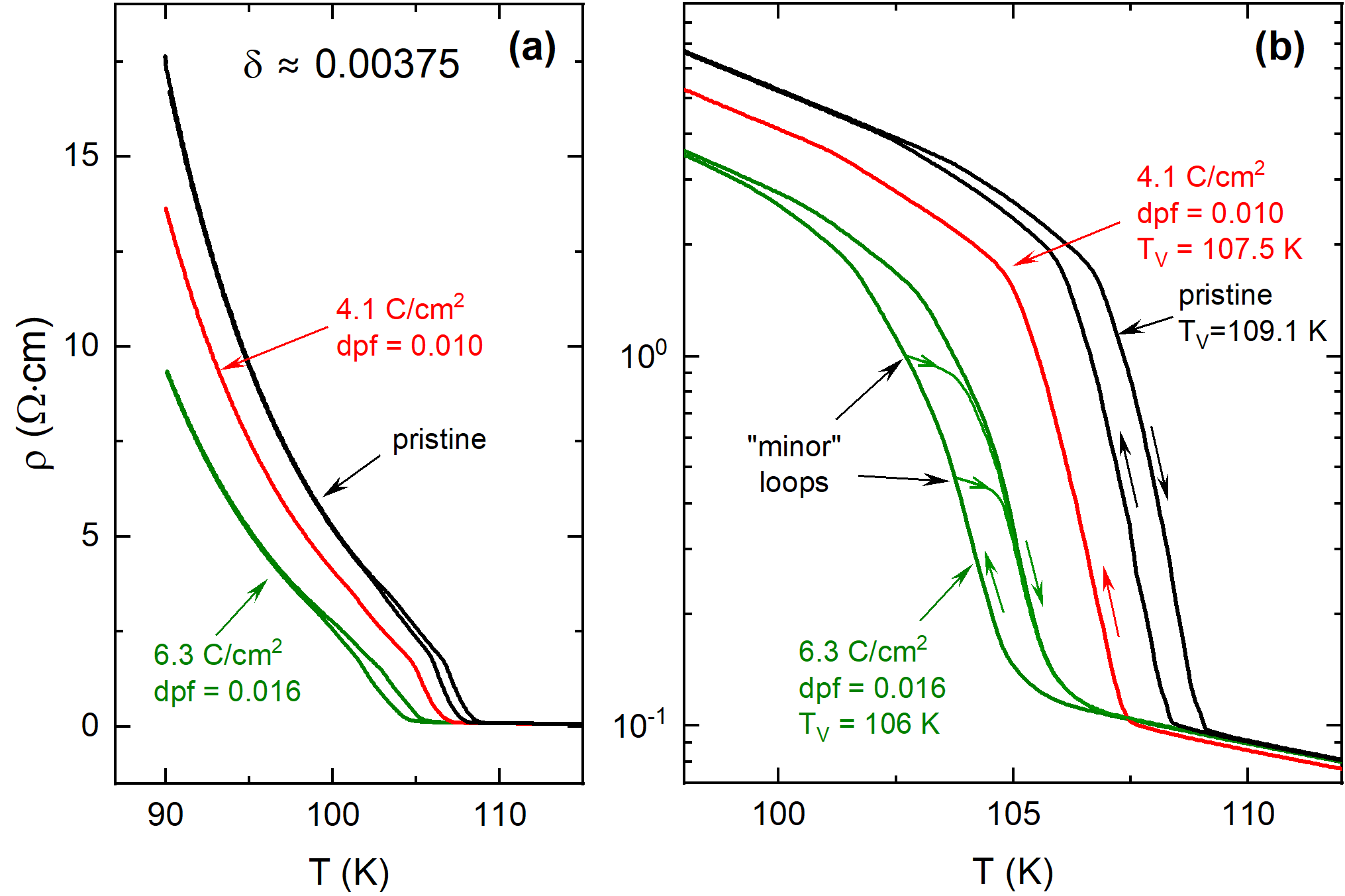}
\centering
\caption{Measurements similar to Fig.\ref{fig:R(T)optimal} performed on an
off-steocheometric magnetite crystal with $T_{V}\approx109$ K. Two
large doses of electron irradiation were applied in this case, 4.1 C/cm$^{2}$ (red curve - cooling
only) of 10 defects per 1000 formula units and 6.3
C/cm$^{2}$ (green curve, cooling and warming) corresponding to 16
defects per 1000 formula units. In addition, zoomed in
6.3 C/cm$^{2}$ curve in panel (b) shows two ``minor'' hysteresis
loops when the sample was cooled to the middle of the transition and
then warmed up showing switching between the two branches. Quite similar
observations may be drawn from this figure as compared to the optimal
crystal data of Fig.\ref{fig:R(T)optimal}.}
\label{fig:R(T)lowTv}
\end{figure}

Let us switch to an off-stoichiometric magnetite crystal with $T_{V}\approx109.1$
K. Figure \ref{fig:R(T)lowTv} shows measurements similar to Fig.\ref{fig:R(T)optimal}.
Two large doses of electron irradiation were applied in this case,
4.1 C/cm$^{2}$
(red curve - cooling only) and 6.3 C/cm$^{2}$ (green curve, cooling and warming). In addition, zoomed in 6.3 C/cm$^{2}$
curve in Fig.\ref{fig:R(T)lowTv}(b) shows two ``minor'' hysteresis
loops when the sample was cooled to the middle of the transition and
then warmed up showing switching between the two branches. This is
very characteristic of an irreversible behavior. The question
is - what is the origin of the hysteretic resistivity? Structural
domains would contribute randomly and it is unclear what kind of domains
is formed in a slightly distorted monoclinic phase. It seems that
a substantial increase of magnetic anisotropy is responsible for such
behavior and the hysteresis is connected to under-cooling and super-heating
associated with the 1st order transition. Introduced point-like disorder
makes the system less perfect and the hysteresis even shrinks
somewhat.

\begin{figure}[tb]
\includegraphics[width=8cm]{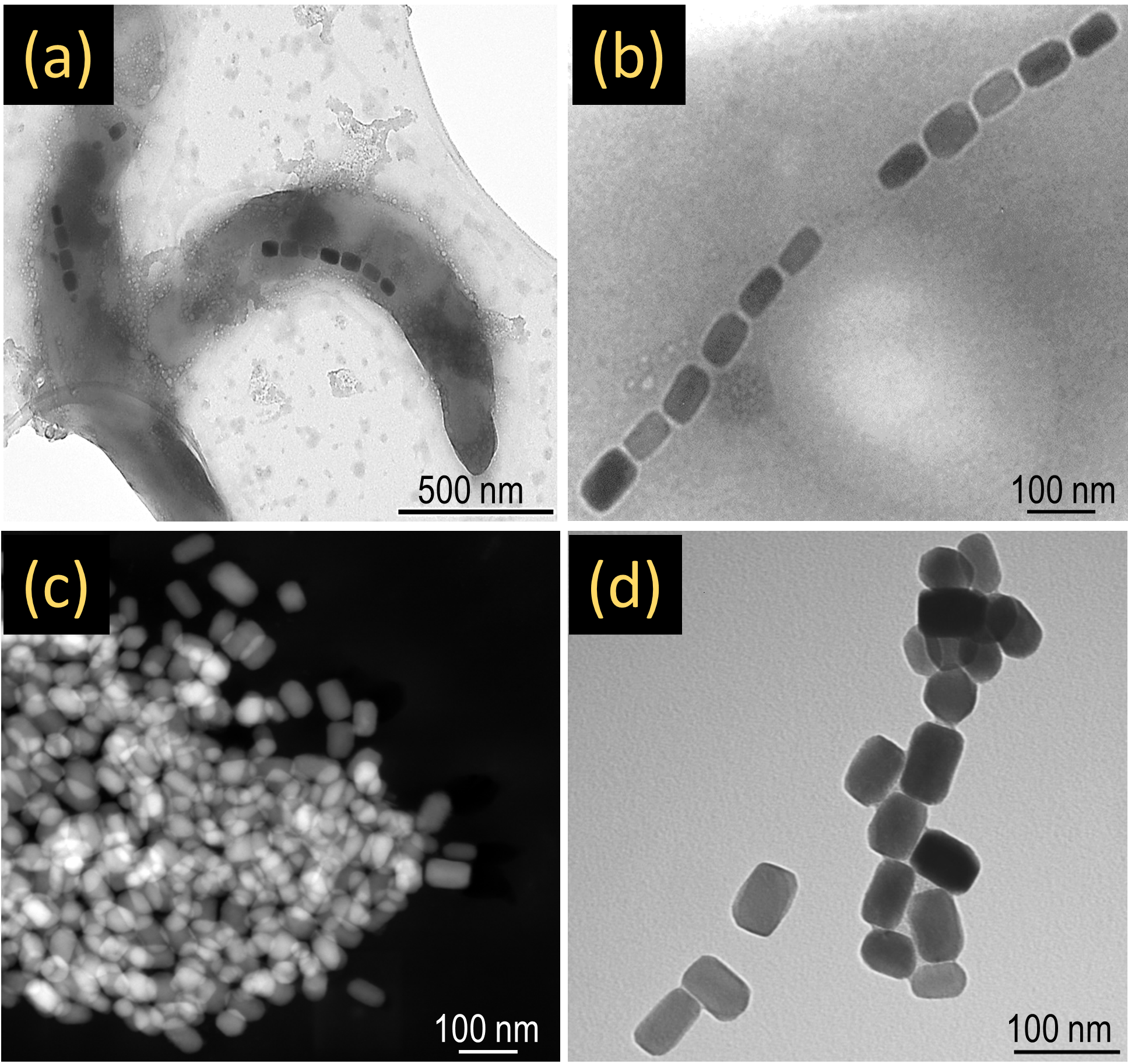}
\centering
\caption{Biogenic magnetite nanocrystals extracted from MV-1 strain of magnetotactic
bacteria. (a) TEM image of the whole bacteria with magnetosome containing
a chain of {[}111{]} oriented head-to-tail magnetite particles inside.
(b) Zoom in on a magnetosome chain. (c) bacteria lised using French
press (d) free magnetite nanoparticles with {[}111{]} direction along
the longer side.}
\label{fig:TEM}
\end{figure}

We now examine biogenic magnetite extracted from magneto-tactic bacteria.
Figure \ref{fig:TEM} shows transmission electron microscopy (TEM)
images of magnetotactic bacteria MV-1 that produce some of the most
perfect nanocrystals of magnetite \citep{DuninBorkowski1998,Dubbels2004,Prozorov2007}.
The extracted magnetite nanocrystals were smeared over the TEM grid
and the grid was encapsulated in a thin aluminum foil. This ``pack''
was measured in \emph{Quantum Design} MPMS before and after irradiation.

\begin{figure}[tb]
\includegraphics[width=8cm]{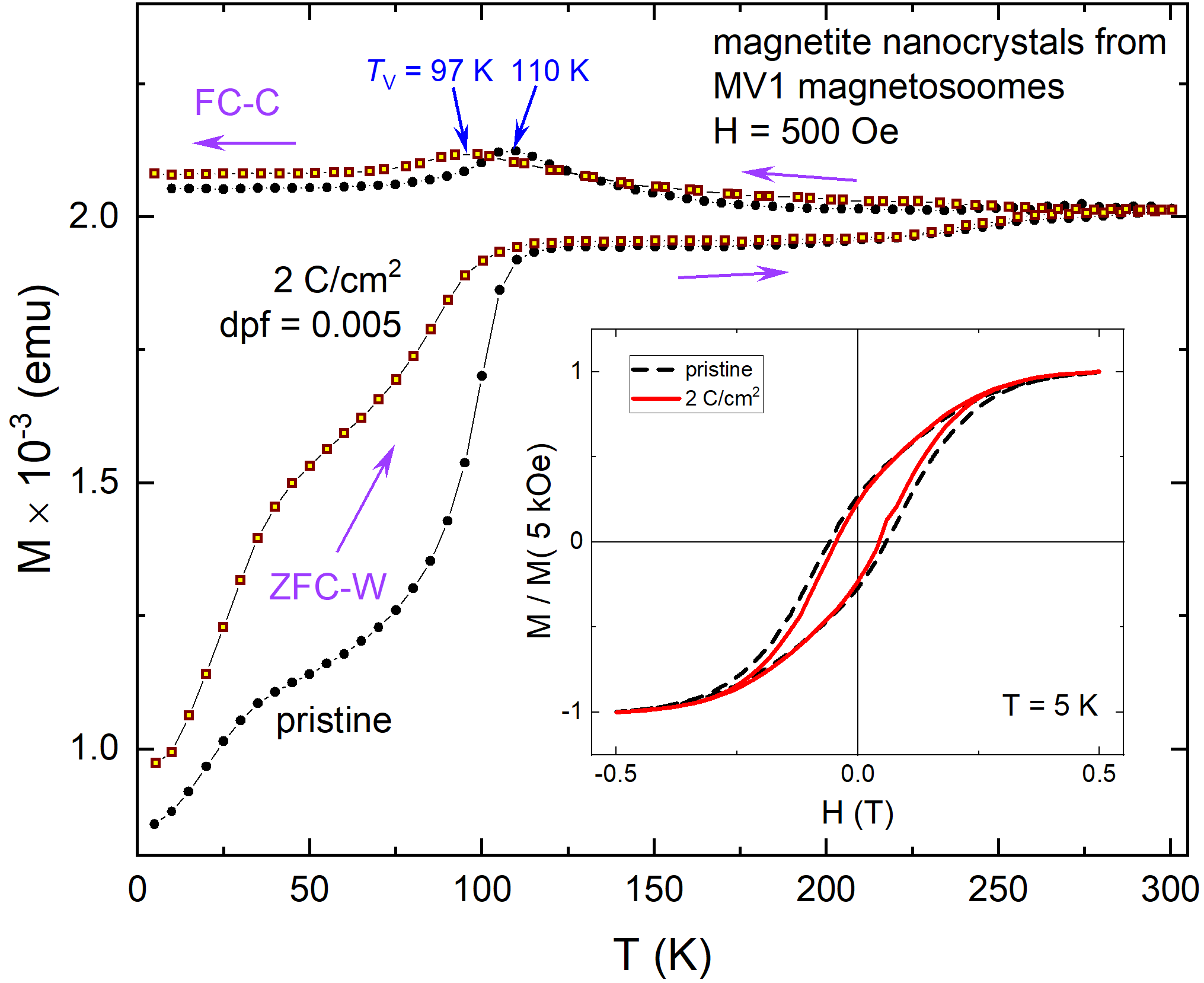}
\centering
\caption{Magnetic measurements of biogenic magnetite. Main panel shows zero-field
cooled (ZFC) measurements, followed by field-cooled (FC) data, both
in a $H=500$ Oe applied magnetic field. Black line is for pristine
sample and red after 2 C/cm$^{2}$ electron-irradiated sample. The
inset shows low-temperature $M(H)$ loops of the same sample at $T=5$
K before (dashed black line) and after (solid red line) irradiation.}
\label{fig:magnetosomes}
\end{figure}

Figure \ref{fig:magnetosomes} shows temperature and field dependent
magnetization of magnetite nanocrystals extracted from the MV-1 magnetotactic
bacteria. Main panel shows magnetization measured on warming after
the sample was cooled to 5 K in zero field, then $H=500$ Oe magnetic
field was applied and the data were collected on warming (this protocol
is abbreviated ZFC-W). After reaching maximum temperature, the measurements
continued on cooling without turning magnetic field off (FC-C protocol).
Pristine sample curve is shown by black filled circles. The curve
after electron irradiation of 2 C/cm$^{2}$ is shown by open red squares.
The inset in Fig.\ref{fig:magnetosomes} shows $T=5$ K $M(H)$ loops
of the same sample, before (dashed black line) and after (solid red
line) irradiation. Notice the significant hysteresis between ZFC and
FC measurements at all temperatures closing only above 250 K. This
means that blocking (collective irreversibility) temperature is somewhere
around room temperature \citep{Prozorov2007,Prozorov1999}. Therefore,
the direction of the magnetic moments in mono-domain nanocrystals
is fixed ($k_{B}T_{V}\ll k_{B}T_{B}$), hence the sharp changes around
110 K are due to the Verwey transition. Similar to our observations
on crystals described above, Verwey temperature shifts to the lower
values and hysteresis becomes somewhat smaller after irradiation.

\begin{figure}[tb]
\includegraphics[width=8cm]{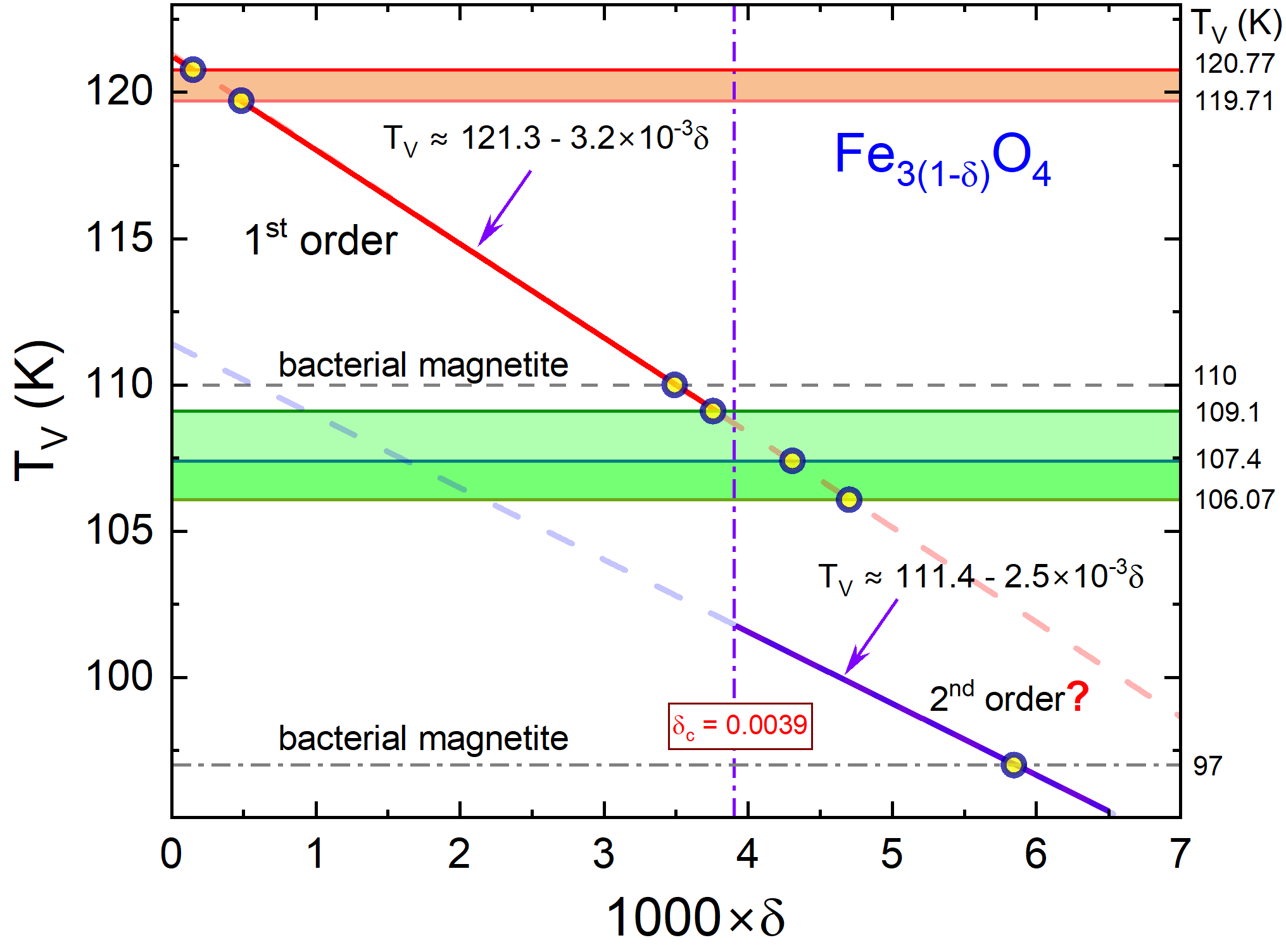} \caption{Verwey transition temperature versus off-stoichiometric parameter,
$\delta$, in Fe$_{3\left(1-\delta\right)}$O$_{4}$. The solid lines
are the fits to the multiple literature data collected on off-stoichiometric
\citep{Aragon1985,Kakol1989,Honig1986,Honig1995,Rozenberg1996,Shepherd1991}
and ion-substituted magnetite adjusted to the parameter $\delta$
\citep{DjegaMariadassou1989,Kakol1992,Brabers1998,Gasparov2009}.
For $\delta\lesssim0.0039$ the transition is of the first order and
is well approximated by $T_{V}=121.3-3.2\times10^{-3}\delta$, whereas
for $\delta\gtrsim0.0039$ the transition was thought to be of the
2nd order and is described by $T_{V}=111.4-2.5\times10^{-3}\delta$.
Our measurements are shown by symbols. Importantly, driven by irradiation,
the off-stoichiometric sample enters continuously the forbidden range
of $T_{V}$ and shows the same highly hysteretic behavior, against
the second order expectations.}
\label{Tv-vs-delta}
\end{figure}

Finally, we examine the evolution of the Verwey transition temperature
with added point-like disorder induced by 2.5 MeV electron irradiation.
Figure \ref{Tv-vs-delta} shows Verwey transition temperature versus
off-stoichiometric parameter $\delta$ in Fe$_{3\left(1-\delta\right)}$O$_{4}$.
The solid lines are the fits to the multiple literature data collected
on off-stoichiometric \citep{Aragon1985,Kakol1989,Honig1986,Honig1995,Rozenberg1996,Shepherd1991}
and ion-substituted magnetite adjusted to the parameter $\delta$
\citep{DjegaMariadassou1989,Kakol1992,Brabers1998,Gasparov2009}.
Until now, it is a general believe that for $\delta\lesssim0.0039$
the transition is of the first order, and can be well approximated
by $T_{V}=121.3-3.2\times10^{-3}\delta$, whereas for $\delta\gtrsim0.0039$
the transition is of the second order and is described by $T_{V}=111.4-2.5\times10^{-3}\delta$.
In Fig. \ref{Tv-vs-delta} our measurements are shown by symbols.
Clearly, driven by disorder induced by the electron irradiation, the
off-stoichiometric sample enters continuously the ``forbidden''
range of $T_{V}$ and shows the same highly hysteretic behavior, against
the second order expectations. The hysteresis has also been observed
in all previous measurements with no explanation provided of how is
this compatible with the 2nd order transition.

\section{Discussion}

Our temperature-dependent resistivity measurements in pristine samples
are fully consistent with the literature data, so we have a reliable
starting point to investigate the effects of electron irradiation.
There are several important features observed in magnetite crystals
and nanocrystals before and after electron irradiation that introduces
artificial controlled point-like disorder.

(1) The increase of the resistivity above $T_{V}$ is counter-intuitive, considering activated character of transport. This suggests that carrier mobility is affected stronger by disorder than carrier density.

(2) The sharpness of the step-like $\rho\left(T\right)$ transition
remains practically unchanged (contrary to the substitutional disorder,
which significantly smears and broadens the transition \citep{Brabers1998}).
All curves remain parallel at the transition showing no broadening
after the irradiation, they just shift.

(3) The under-cooling / super-heating hysteresis becomes a little
bit smaller after irradiation.

(4) Resistivity is not metallic at all temperatures exhibiting semiconductor-like
Arrhenius activated behavior with characteristic energy barriers around
$U=200$ meV below $T_{V}$ and $60$ meV above. These values are
similar to the reported in literature \citep{Banerjee2020}.

(5) These barrier values remain practically unchanged even after substantial
increase of the disorder.

(6) The Verwey transition temperature in the irradiated sample shifts
by a significant amount of $\Delta T_{V}/T_{V}^{0}\approx0.48\%$
per 1 C/cm$^{2}$ (6.24 $\times$ 10$^{18}$ electrons/cm$^{2}$)
in the optimal sample, and $0.45\%$ per 1 C/cm$^{2}$ in the lower
$T_{V}$ sample. These are quite similar values despite the large
difference in $T_{V}$ and the actual maximum doses acquired, 3.76
C/cm$^{2}$ and 6.3 C/cm$^{2}$, respectively.

(7) The Verwey transition of the lower-$T_{V}$ sample in pristine
state is at the border of an alleged cross-over from the 1st order
to the 2nd order transition. The irradiation, however, pushes Verwey
temperature continuously to lower values without broadening or smearing
and with a substantial hysteresis. This, combined with the original
Honig's observations of such hysteresis means that the discontinuity
in the $T_{V}\left(\delta\right)$ or $T_{V}\left(\mathrm{doping}\right)$
is most likely due to sample inhomogeneity and phase separation rather
than the intrinsic behavior.

(8) Irradiated samples can be annealed toward the pristine state showing
the behavior expected for the lower-dose irradiation.

(9) Magnetite nanocrystals extracted from MV-1 strain of magnetotactic
bacteria show behavior qualitatively similar to the bulk crystals.
The Verwey temperature is suppressed and hysteresis becomes smaller.
Of course, it was only possible to measure magnetization, not electric transport.

Is it possible to understand all those features from a single point
of view? If the Verwey transition was driven by Fe$^{2+}$/Fe$^{3+}$
ordering and/or electrons/polarons localization, it would be significantly
smeared by the additional randomly distributed defects. Also, resistivity
would increase at all temperatures. The same arguments can be used
against the Wigner glass to crystal transition. In our view, the opening of the band-gap the most plausible explanation
of the Verwey transition. Additional disorder, at our low concentrations,
does not affect the energy gap itself, but could form an impurity
(disorder) band close to the conduction band. Additionally, it would
certainly affect/increase the bandwidth making the band edge ``fuzzy''.
According to the theory, this would lower the effective activation
barrier \citep{Brabers2000}, thus leading to the reduction of $T_{V}$.
Furthermore, the increase of $\rho$ above $T_{V},$ and the decrease
below is most likely due to the magnetic component of the problem.
At least some irradiation-induced defects may become ``magnetic''
scatterers when both the defect's and scattered electron's spins flip.
This would be effective in the magnetically mildly anisotropic cubic
phase above $T_{V}$ but then become less significant in the higher
anisotropy monoclinic phase. However, the reason that resistivity
decreases in the low-temperature phase after irradiation is the same
broadening of the bandwidth lowering the effective barrier, hence
leading to more carriers in the conduction band. Apparently this effect
dominates the opposed trend of increased density of scattering centers.

\section{Conclusions}

To summarize, a 2.5 MeV low-temperature (20 K) electron irradiation
was used to induce point-like defects in magnetite crystals and biogenic
nanocrystals. The Verwey transition shifts to the lower temperatures
regardless of the $T_{V}$ itself, approximately at the rate of $\Delta T_{V}/T_{V}^{0}\approx0.5\%$
per 1 C/cm$^{2}$ (6.24 $\times$ 10$^{18}$ electrons/cm$^{2}$).
The transition itself remains sharp without any sign of smearing or
broadening. In $T_{V}=109.1$ K sample, in the pristine state, it
is right at the ``border'' of the alleged 1st-to-2nd order transition.
Upon irradiation, the resistivity curve parallel-shifts to what so far was believed to be a  "forbidden" range for the Vervey's $T_V$. Notably, the transition itself remains sharp and hysteretic arguing against the intrinsic mechanism of the \emph{apparent} 2nd order
transition. The wealth of obtained results can be explained within
the bandwidth bandgap theory with a smaller gap, around 60 meV above
$T_{V}$, and a larger gap around 200 meV below $T_{V}$. We hope
that the obtained data will motivate further work on the mechanisms
of the fascinating Verwey transition in magnetite.

\acknowledgments
We are grateful to Jurgen M. Honig (passed away on November 4, 2022) and to Robert J. McQueeney for illuminating discussions and for providing high-quality single crystals of magnetite; and to Dennis A. Bazylinski for valuable discussions and providing biogenic nanocrystals form the MV-1 magnetotactic bacteria.

This research was supported by the U.S. Department of Energy (DOE), Office of Science, Basic Energy Sciences, Materials Science and Engineering Division. Ames Laboratory is operated for the U.S. DOE by Iowa State University under contract DE-AC02-07CH11358. We thank the SIRIUS team for running electron irradiation at \'{E}cole Polytechnique supported by the EMIR\&A (the French Federation of Accelerators for Irradiation and Analysis of Molecules and Materials) user proposals 15-2463, 16-0640 and 17-8487.


%

\end{document}